\def\mode{b }  

\documentstyle[12pt]{article}

\newif\ifprelim
\def\paperdate{May 31, 1995}

\def\baselinestretch{1.2}
\def\unlock{\catcode`@=11 }
\def\lock{\catcode`@=12 }

\def\marginnote#1{}
\let\ju=\marginnote
\def\mybiblabel#1{#1\hfil}
\unlock
\def\@bibitem#1{\def\@blbl{#1}\item\if@filesw \immediate\write\@auxout
        {\string\bibcite{#1}{\the\value{\@listctr}}}\fi\ignorespaces}
\lock
\def\blackfonts{
                \font\blackboard=msbm10 scaled\magstep1
                \font\blackboards=msbm8
                \font\blackboardss=msbm6}

\def\littlemode{l }
\def\draftmode{d }
\def\Draftmode{D }

\def\Newif#1{\expandafter\ifx\csname#1\endcsname\relax
 \csname newif\expandafter\endcsname\csname#1\endcsname\fi}
\Newif{ifprelim}

\ifprelim\let\paperdate\today\fi

\ifx\mode\undefined
\def\bigpage{
        \textheight 22.5 cm
        \topmargin -.75 cm
        \textwidth 16cm
        \marginparwidth 54pt
        \marginparsep 8pt
        \oddsidemargin 0 in
        \evensidemargin 0 in
}
\def\doublepage{                           
        \parskip 6 pt
        \parindent 2em
        \twocolumn
        
        \let\small\relax
        \let\sl\it
        \sloppy
        \voffset=-2.54truecm
        \hoffset=-1.54truecm
        \flushbottom
        \leftmargini 2em
        \leftmarginv .5em
        \leftmarginvi .5em
        \marginparwidth 48pt
        \marginparsep 8pt
        \columnsep 2truecm
        \textwidth 25.4truecm
        \textheight 17truecm
        \oddsidemargin .18truein
        \evensidemargin .17truein
}
\def\draftpage{
        \textheight 19.86075cm
        \topmargin -.75 cm
        \textwidth 13.66667cm
        \oddsidemargin  -.25 cm
        \evensidemargin -.25 cm
        \marginparwidth 72pt
        \marginparsep 8pt
}
\def\smallblack{
        \def\blackfonts{
                \font\blackboard=msbm10
                \font\blackboards=msbm7
                \font\blackboardss=msbm5
        }
}
\def\marginnotes{
        \def\draftmarginnote##1{\marginpar{\raggedright\scriptsize\tt ##1}}
        \let\marginnote=\draftmarginnote
        \let\ju=\marginnote
}
\unlock

\def\draftnotice{
        \def\@oddhead{\phantom{\em \standardtime~\today} \hfil
                \smash{\Large\em DRAFT} \hfil
                \em \standardtime~\today}
        \let\@evenhead\@oddhead
        \def\ps@plain{\let\@mkboth\@gobbletwo
                \def\@oddfoot{\hfil 
                \thepage 
                \hfil}
                \let\@evenfoot\@oddfoot}
        \def\ps@empty{\let\@mkboth\@gobbletwo
                \def\@oddfoot{\hfil \smash{\Large\em DRAFT} \hfil}
                \let\@evenfoot\@oddfoot}
        \pagestyle{plain}
        \draftspecial
}
\def\eqnlabels{
        \def\draftlabel##1{{\@bsphack\if@filesw {\let\thepage\relax
           \xdef\@gtempa{\write\@auxout{\string
              \newlabel{##1}{{\@currentlabel}{\thepage}}}}}\@gtempa
           \if@nobreak \ifvmode\nobreak\fi\fi\fi\@esphack}
                \gdef\@eqnlabel{##1}}
        \def\@eqnlabel{}
        \def\@vacuum{}
        \let\label=\draftlabel
        \def\@eqnnum{(\theequation)\rlap{\kern\marginparsep\tt\@eqnlabel}%
                \global\let\@eqnlabel\@vacuum}
}
\def\draftbib{
        \def\mybiblabel##1{\llap{\scriptsize\tt \@blbl\ }##1\hfil}
        \def\@lbibitem[##1]##2{%
                \def\@blbl{##2}\item[\@biblabel{##1}\hfill]\if@filesw
                {\def\protect##1{\string ##1\space}\immediate
                \write\@auxout{\string\bibcite{##2}{##1}}}\fi\ignorespaces}
        \def\@bibitem##1{\def\@blbl{##1}\item\if@filesw
                \immediate\write\@auxout
                {\string\bibcite{##1}{\the\value{\@listctr}}}\fi\ignorespaces}
        }
\lock
\let\draftspecial=\draftport

\ifx\mode\littlemode
        \ifprelim
                \let\draftspecial=\draftland
                \draftnotice
        \else
        \fi
        \unlock
        \input art10.sty
        \lock
        \def\baselinestretch{1.3}
        \smallblack
        \doublepage
\else\ifx\mode\draftmode
        \draftpage
        \draftnotice
        \marginnotes
        \eqnlabels
        \draftbib
\else\ifx\mode\Draftmode
        \bigpage
        \advance\oddsidemargin  -5 mm
        \advance\evensidemargin -5 mm
        \draftnotice
        \marginnotes
        \eqnlabels
        \draftbib
\else
        \ifprelim
                \draftnotice
        \else
        \fi
        \bigpage

\fi
\fi
\fi

\batchmode
        \newfont{\footbbbfont}{msbm10}
\errorstopmode

\newif\ifamsf\amsftrue
        \ifx\footbbbfont\nullfont
        \amsffalse
\fi

\ifamsf
        \blackfonts
        \newfam\black
        \textfont\black=\blackboard
        \scriptfont\black=\blackboards
        \scriptscriptfont\black=\blackboardss

        \def\CC{{\fam\black\relax C}}

\else            

        \def\CC{{C \n{11} C}}

\fi

\newcount\hour
\newcount\minute
\newtoks\amorpm
\hour=\time\divide\hour by 60
\minute=\time{\multiply\hour by 60 \global\advance\minute by-\hour}
\def\standardtime{{\ifnum\hour<12 \global\amorpm={AM}%
        \else\global\amorpm={PM}\advance\hour by-12 \fi
        \ifnum\hour=0 \hour=12 \fi
        \number\hour:\ifnum\minute<10 0\fi\number\minute\the\amorpm}}
\def\militarytime{\number\hour:\ifnum\minute<10 0\fi\number\minute}

\newif\ifepsfloaded
\newif\iffigureexists

\ifeof 1 \epsfloadedfalse \else \epsfloadedtrue \fi
\ifepsfloaded \input epsf \fi

\def\checkex#1 {\relax
    \openin 1 #1
    \ifeof 1 \figureexistsfalse
    \else \figureexiststrue
    \fi \closein 1 }

\def\cpsbox#1#2{
        \ifepsfloaded
                \checkex #2
                \iffigureexists
                        \immediate\write16{(#2)}
                        \setlength{\epsfxsize}{#1}
                        \centerline{\epsfbox{#2}}
                \else
                        \immediate\write16{(#2 NOT FOUND!)}
                        \vbox to 2in{\hbox to #1 {\hss} \vss}
                \fi
        \else
                \immediate\write16{(NOT inputting #2; no epsf.tex)}
                \vbox to 2in{\hbox to #1 {\hss} \vss}
        \fi}
\unlock
\def\@citex[#1]#2{%
\if@filesw \immediate \write \@auxout {\string \citation {#2}}\fi
\@tempcntb\m@ne \let\@h@ld\relax \def\@citea{}%
\@cite{%
  \@for \@citeb:=#2\do {%
    \@ifundefined {b@\@citeb}%
      {\@h@ld\@citea\@tempcntb\m@ne{\bf ?}%
      \@warning {Citation `\@citeb ' on page \thepage \space undefined}}%
      {\@tempcnta\@tempcntb \advance\@tempcnta\@ne%
      \@tempcntb\number\csname b@\@citeb \endcsname \relax%
      \ifnum\@tempcnta=\@tempcntb 
        \ifx\@h@ld\relax%
          \edef \@h@ld{\@citea\csname b@\@citeb\endcsname}%
        \else%
          \edef\@h@ld{\ifmmode{-}\else--\fi\csname b@\@citeb\endcsname}%
        \fi%
      \else
        \@h@ld\@citea\csname b@\@citeb \endcsname%
        \let\@h@ld\relax%
      \fi}%
    \def\@citea{,\penalty\@highpenalty\,}%
  }\@h@ld
}{#1}}

\def\@citeb#1#2{{[#1]\if@tempswa , #2\fi}}
\def\@citeu#1#2{{$^{#1}$\if@tempswa , #2\fi }}
\def\@citep#1#2{{#1\if@tempswa , #2\fi}}

\def\bcites{         
        \unlock
        \let\@cite=\@citeb
        \lock
}

\def\upcites{         
        \unlock
        \let\@cite=\@citeu
        \lock
}

\def\plaincites{      
        \unlock
        \let\@cite=\@citep
        \lock
}

\let\@cite=\@citeb              

\def\refname{References}        

\def\thebibliography#1{\section*{\refname\@mkboth
  {\uppercase{\refname}}{\uppercase{\refname}}}\list
  {\@biblabel{\arabic{enumiv}}}{\settowidth\labelwidth{\@biblabel{#1}}%
    \let\makelabel\mybiblabel\leftmargin\labelwidth
    \advance\leftmargin\labelsep
    \usecounter{enumiv}%
    \let\p@enumiv\@empty
    \def\theenumiv{\arabic{enumiv}}}%
    \def\newblock{\hskip .11em plus.33em minus.07em}%
    \sloppy\clubpenalty4000\widowpenalty4000
    \sfcode`\.=1000\relax}

\def\@noitemerr{\@warning{Something's wrong--perhaps a missing
\string\item}\@ehc}

\def\sections{\unlock
\def\theequation{\thesection.\arabic{equation}}
\@addtoreset{equation}{section}
\@addtoreset{footnote}{section}
\lock
}

\def\footnotesections{\unlock
\@addtoreset{footnote}{section}
\lock
}

\def\subsections{\unlock
\def\theequation{\thesubsection.\arabic{equation}}
\@addtoreset{equation}{subsection}
\@addtoreset{footnote}{subsection}
\lock
}

\def\emptyliststuff#1{\par \penalty -500 
        \noindent $\bullet$ {\bf #1}}
\def\liststuff#1{\par \penalty -500 
        \noindent $\bullet$ {\bf #1}\\*}

\lock

\def\cf{{\it c.f.\/}}

\def\ibid{{\it ibid.\/}}
\def\ie{\hbox{\it i.e.\/}}

\def\CS{Chern--Simons}

\def\YM{Yang--Mills}

\def\noj#1,#2,{{\bf #1} (19#2)\ }
\def\jou#1#2,#3,{{\em #1\/ }{\bf #2} (19#3)\ }
\def\hep{hep-th/}
\def\ann#1,#2,{{\em Ann.\ Physics\/ }{\bf #1} (19#2)\ }
\def\annmath#1,#2,{{\em Ann.\ Math\/ }{\bf #1} (19#2)\ }
\def\cmp#1,#2,{{\em Comm.\ Math.\ Phys.\/ }{\bf #1} (19#2)\ }
\def\cq#1,#2,{{\em Class.\ Quantum Grav.\/ }{\bf #1} (19#2)\ }
\def\cqg#1,#2,{{\em Class.\ Quantum Grav.\/ }{\bf #1} (19#2)\ }
\def\ijmp#1,#2,{{\em Int.\ J.\ Mod.\ Phys.\/ }{\bf A#1} (19#2)\ }
\def\jdiff#1,#2,{{\em J.\ Diff.\ Geom.\/ }{\bf #1} (19#2)\ }
\def\jmp#1,#2,{{\em J.\ Math.\ Phys.\/ }{\bf #1} (19#2)\ }
\def\jp#1,#2,{{\em J.\ Phys.\/ }{\bf A#1} (19#2)\ }
\def\grg#1,#2,{{\em Gen.\ Rel.\ Grav.\/ }{\bf #1} (19#2)\ }
\def\lmp#1,#2,{{\em L.\ Math.\ Phys.\/ }{\bf #1} (19#2)\ }
\def\mpl#1,#2,{{\em Mod.\ Phys.\ Lett.\/ }{\bf A#1} (19#2)\ }
\def\nc#1,#2,{{\em Nuovo Cim.\/ }{\bf #1} (19#2)\ }
\def\np#1,#2,{{\em Nucl.\ Phys.\/ }{\bf B#1} (19#2)\ }
\def\philos#1,#2,{{\em Philos.\ Trans.\ R.\ Soc.\ Lond.\/ }{\bf A#1} (19#2)\ }
\def\procroysoc#1,#2,{{\em Proc.\ Roy.\ Soc.\ Lond.\/ }{\bf A#1} (19#2)\ }
\def\pl#1,#2,{{\em Phys.\ Lett.\/ }{\bf #1B} (19#2)\ }
\def\pla#1,#2,{{\em Phys.\ Lett.\/ }{\bf #1A} (19#2)\ }
\def\pr#1,#2,{{\em Phys.\ Rev.\/ }{\bf #1} (19#2)\ }
\def\prd#1,#2,{{\em Phys.\ Rev.\/ }{\bf D#1} (19#2)\ }
\def\prl#1,#2,{{\em Phys.\ Rev.\ Lett.\/ }{\bf #1} (19#2)\ }
\def\prp#1,#2,{{\em Phys.\ Rept.\/ }{\bf #1C} (19#2)\ }
\def\ptp#1,#2,{{\em Prog.\ Theor.\ Phys.\/ }{\bf #1} (19#2)\ }
\def\ptpsup#1,#2,{{\em Prog.\ Theor.\ Phys.\/ Suppl.\/ }{\bf #1} (19#2)\ }
\def\rmp#1,#2,{{\em Rev.\ Mod.\ Phys.\/ }{\bf #1} (19#2)\ }
\def\yadfiz#1,#2,#3[#4,#5]{{\em Yad.\ Fiz.\/ }{\bf #1} (19#2) #3
[{\em Sov.\ J.\ Nucl.\ Phys.\/ }{\bf #4} (19#2) #5]}
\def\zh#1,#2,#3[#4,#5]{{\em Pis'ma\ Zh.\ Exp.\ Theor.\ Fiz.\/ }{\bf #1}
(19#2) #3 [{\em Sov.\ Phys.\ JETP\/ }{\bf #4} (19#2) #5]}

\def\eq#1{.~(\ref{#1})}
\def\noeq#1{(\ref{#1})}
\def\beq{\begin{equation}}
\def\eeq{\end{equation}}
\def\beqar{\begin{eqnarray}}
\def\eeqar{\end{eqnarray}}

\def\beqal{\begin{equation}\begin{eqalign}}
\def\eeqal{\end{eqalign}\end{equation}}
\def\beqalno{\begin{eqalignno}}
\def\eeqalno{\end{eqalignno}}
\def\beqaltwo{\begin{eqaligntwo*}}
\def\eeqaltwo{\end{eqaligntwo*}}
\def\bsubeq{\begin{subequations}}
\def\esubeq{\end{subequations}}

\def\u#1{{}^{#1}}

\def\p#1{\mskip#1mu}
\def\n#1{\mskip-#1mu}
\def\stop{\p6.}
\def\comma{\p6,}
\def\semi{\p6;}

\def\to{\rightarrow}

\def\longlongrightarrow{\relbar\joinrel\relbar\joinrel\rightarrow}

\def\onarrow#1{\mathrel{\mathop{\longrightarrow}\limits^{#1}}}
\def\onArrow#1{\mathrel{\mathop{\longlongrightarrow}\limits^{#1}}}

\def\nfrac#1#2{{\displaystyle{\vphantom1\smash{\lower.5ex\hbox{\small$#1$}}%
        \over\vphantom1\smash{\raise.25ex\hbox{\small$#2$}}}}}

\def\vev#1{\left\langle #1 \right\rangle}

\def\norm#1{\left|\left| #1 \right|\right|}
\def\com#1#2{ \left [ #1 \, , #2 \, \right ] }
\def\anticom#1#2{ \left \{ #1 \, , #2 \, \right \} }

\def\pa{\partial}

\def\Tr{{\rm Tr}}
\def\l:{\mathopen{:}\,}
\def\r:{\,\mathclose{:}}

\def\det{\mathop{\rm det}\nolimits}

\unlock
\def\@versim#1#2{\smash{\lower0.5ex\vbox{\baselineskip\z@skip\lineskip\z@skip
        \lineskiplimit\z@\ialign{$\m@th#1\hfil##\hfil$\crcr#2\crcr\sim\crcr}}}}
\def\ltsim{\mathrel{\mathpalette\@versim<}}
\def\gtsim{\mathrel{\mathpalette\@versim>}}
\lock

\def\bop#1{\setbox0=\hbox{$#1M$}\mkern1.5mu
        \vbox{\hrule height0pt depth.04\ht0
        \hbox{\vrule width.04\ht0 height.9\ht0 \kern.9\ht0
        \vrule width.04\ht0}\hrule height.04\ht0}\mkern1.5mu}
\def\Box{{\mathpalette\bop{}}}

\def\uone{$U(1)$}
\def\utwo{$U(2)$}

\def\sutwo{$SU(2)$}
\def\sufour{$SU(4)$}

\def\sothree{$SO(3)$}
\def\sofour{$SO(4)$}

\def\ntwo{\hbox{$N=2$}}
\def\nfour{\hbox{$N=4$}}

\def\twoi{2 \p2 i}
\def\fouri{4 \p2 i}

\def\warboxp#1{\setbox0=\hbox{$#1M$}\mkern1.5mu
        \vbox{\hrule height0pt depth.04\ht0
        \hbox{\vrule width.04\ht0 height.9\ht0 \kern.9\ht0
        \vrule width.04\ht0}\hrule height.04\ht0}\mkern1.5mu}
\def\warbox{{\mathpalette\warboxp{}}}                        

\def\boxes#1{
        \newcount\num
        \num=1
        \newdimen\downsy
        \downsy=-1.5ex
        \mkern-3.5mu
        \warbox
        \loop
        \ifnum\num<#1
        \llap{\raise\num\downsy\hbox{$\warbox$}}
        \advance\num by1
        \repeat}
\def\boxup#1#2{\newcount\numup
        \numup=#1
        \newdimen\upsy
        \upsy=.75ex
        \mkern3.5mu
        \raise -.25ex\hbox{{\raise\numup\upsy\hbox{$#2$}}}}

\def\cala{{\cal A}}

\def\calc{{\cal C}}
\def\cald{{\cal D}}

\def\calf{{\cal F}}
\def\calg{{\cal G}}

\def\call{{\cal L}}
\def\calm{{\cal M}}

\def\cals{{\cal S}}

\def\abar{{\bar a}}

\def\Abar{{\bar A}}

\def\Dbar{{\bar D}}

\def\Fbar{{\bar F}}

\def\Qbar{{\bar Q}}

\def\etabar{{\bar \eta}}

\def\psibar{{\bar \psi}}

\unlock
\def\section{\@startsection {section}{1}{\z@}{3.ex plus 1ex minus
 .2ex}{2.ex plus .2ex minus 1ex}{\large\bf}}
\def\subsection{\@startsection{subsection}{2}{\z@}{2.75ex plus 1ex minus
 .2ex}{1.5ex plus .2ex minus 1ex}{\bf}}
\if@twocolumn
                
                \def\abstractindent{1.5 cm}
\else
                
                \def\abstractindent{.75 cm}
\fi

\def\appendix{{\newpage\section*{Appendices}}\let\appendix\section%
        {\setcounter{section}{0}
        \gdef\thesection{\Alph{section}}}\section}

\def\abstract{
        \vfill\begin{center}
        {\bf Abstract}
        \end{center}
        \advance\leftskip\abstractindent
        \advance\rightskip\abstractindent
}

\lock

\unlock

\newif\if@defeqnsw \@defeqnswtrue

\def\eqnarray{\stepcounter{equation}\let\@currentlabel=\theequation
\if@defeqnsw\global\@eqnswtrue\else\global\@eqnswfalse\fi
\global\@eqnswtrue
\tabskip\@centering\let\\=\@eqncr
$$\halign to \displaywidth\bgroup\hfil\global\@eqcnt\z@
  $\displaystyle\tabskip\z@{##}$&\global\@eqcnt\@ne
  \hfil$\displaystyle{{}##{}}$\hfil
  &\global\@eqcnt\tw@ $\displaystyle{##}$\hfil
  \tabskip\@centering&\llap{##}\tabskip\z@\cr}

\def\yesnumber{\global\@eqnswtrue}

\def\@@eqncr{\let\@tempa\relax\global\advance\@eqcnt by \@ne
    \ifcase\@eqcnt \def\@tempa{& & & &}\or \def\@tempa{& & &}\or
     \def\@tempa{& &}\or \def\@tempa{&}\else\fi
     \@tempa \if@eqnsw\@eqnnum\stepcounter{equation}\fi
     \if@defeqnsw\global\@eqnswtrue\else\global\@eqnswfalse\fi
     \global\@eqcnt\z@\cr}

\def\@eqnacr{{\ifnum0=`}\fi\@ifstar{\@yeqnacr}{\@yeqnacr}}

\def\@yeqnacr{\@ifnextchar [{\@xeqnacr}{\@xeqnacr[\z@]}}

\def\@xeqnacr[#1]{\ifnum0=`{\fi}\cr \noalign{\vskip\jot\vskip #1\relax}}

\def\eqalign{\null\,\vcenter\bgroup\openup1\jot \m@th \let\\=\@eqnacr
\ialign\bgroup\strut
\hfil$\displaystyle{##}$&$\displaystyle{{}##}$\hfil\crcr}
\def\endeqalign{\crcr\egroup\egroup\,}

\def\cases{\left\{\,\vcenter\bgroup\normalbaselines\m@th \let\\=\@eqnacr
    \ialign\bgroup$##\hfil$&\quad##\hfil\crcr}
\def\endcases{\crcr\egroup\egroup\right.}

\def\eqalignno{\stepcounter{equation}\let\@currentlabel=\theequation
\if@defeqnsw\global\@eqnswtrue\else\global\@eqnswfalse\fi
\let\\=\@eqncr
$$\displ@y \tabskip\@centering \halign to \displaywidth\bgroup
  \global\@eqcnt\@ne\hfil
  $\@lign\displaystyle{##}$\tabskip\z@skip&\global\@eqcnt\tw@
  $\@lign\displaystyle{{}##}$\hfil\tabskip\@centering&
  \llap{\@lign##}\tabskip\z@skip\crcr}

\def\endeqalignno{\@@eqncr\egroup
      \global\advance\c@equation\m@ne$$\global\@ignoretrue}

\@namedef{eqalignno*}{\@defeqnswfalse\eqalignno}
\@namedef{endeqalignno*}{\endeqalignno}

\def\eqaligntwo{\stepcounter{equation}\let\@currentlabel=\theequation
\if@defeqnsw\global\@eqnswtrue\else\global\@eqnswfalse\fi
\let\\=\@eqncr
$$\displ@y \tabskip\@centering \halign to \displaywidth\bgroup
  \global\@eqcnt\m@ne\hfil
  $\@lign\displaystyle{##}$\tabskip\z@skip&\global\@eqcnt\z@
  $\@lign\displaystyle{{}##}$\hfil\qquad&\global\@eqcnt\@ne
  \hfil$\@lign\displaystyle{##}$&\global\@eqcnt\tw@
  $\@lign\displaystyle{{}##}$\hfil\tabskip\@centering&
  \llap{\@lign##}\tabskip\z@skip\crcr}

\def\endeqaligntwo{\@@eqncr\egroup
      \global\advance\c@equation\m@ne$$\global\@ignoretrue}

\@namedef{eqaligntwo*}{\@defeqnswfalse\eqaligntwo}
\@namedef{endeqaligntwo*}{\endeqaligntwo}

\newtoks\@stequation

\def\subequations{\refstepcounter{equation}%
  \edef\@savedequation{\the\c@equation}%
  \@stequation=\expandafter{\theequation}
  \edef\@savedtheequation{\the\@stequation}
  \edef\oldtheequation{\theequation}%
  \setcounter{equation}{0}%
  \def\theequation{\oldtheequation\alph{equation}}}

\def\endsubequations{%
  \setcounter{equation}{\@savedequation}%
  \@stequation=\expandafter{\@savedtheequation}%
  \edef\theequation{\the\@stequation}%
  \global\@ignoretrue}

\def\big#1{{\hbox{$\left#1\vcenter to1.428\ht\strutbox{}\right.\n@space$}}}
\def\Big#1{{\hbox{$\left#1\vcenter to2.142\ht\strutbox{}\right.\n@space$}}}
\def\bigg#1{{\hbox{$\left#1\vcenter to2.857\ht\strutbox{}\right.\n@space$}}}
\def\Bigg#1{{\hbox{$\left#1\vcenter to3.571\ht\strutbox{}\right.\n@space$}}}

\lock

\def\com#1#2{ \left [ \vphantom{\tilde \psi^\mu} #1 \, , #2 \, \right ] }
\def\anticom#1#2{ \left \{ \vphantom{\tilde \psi^\mu} #1 \, , #2 \, \right \} }
\def\norm#1{\left|\left| \, \vphantom{\tilde \psi^\mu} #1 \, \right|\right|}

\def\A{\cala}
\def\Abar{\widetilde \cala}
\def\D{\cald}
\def\Dbar{\widetilde \cald}
\def\F{\calf}
\def\Fbar{\widetilde \calf}
\def\chistar{\u* \n1 \chi}
\def\Fstar{\u* \n2 \calf}
\def\Fstarbar{\u* \n2 \widetilde \calf}
\def\munu{{\mu\nu}}
\def\V{V}
\def\Q{\mathop Q }
\def\Qbar{\mathop {\widetilde Q}}
\def\etabar{{\widetilde \eta}}

\def\psibar{{\widetilde \psi}}

\def\four{\hbox{\bf 4}}
\def\sutwotwoone{$SU(2) \otimes SU(2)\otimes U(1)$}
\def\gc{G^{\p2\CC}}
\def\g{G}
\def\pim{\pi_1}
\def\H#1{H_{\n5\A}^{\p1#1}}

\def\boxa{\Box_{\n1 \A}}
\def\boxamunu{\Box_{\n2 \A}^{\p1\munu}}
\def\boxamunuT{\Box_{\n2 \A}^{\p1 T \p1 \munu}}

\sections
\footnotesections

\begin{document}
\begin{titlepage}

\noindent\paperdate\hfill    TAUP--2247--95\\
\null\hfill          hep-th/9506002

\vskip 1.5 cm

\begin{center}

{\large \bf
The other topological twisting of $N=4$ \YM
}

\vskip 1 cm
{
Neil Marcus\footnote{
  E--Mail: \hbox{NEIL@POST.TAU.AC.IL}.\\
  Work supported in part by the US-Israel Binational Science
  Foundation, the German-Israeli Foundation for Scientific Research and
  Development and the Israel Academy of Science.
  }
}
\vskip 0.3 cm

{\sl
  School of Physics and Astronomy\\Raymond and Beverly Sackler Faculty
  of Exact Sciences\\Tel-Aviv University\\Ramat Aviv, Tel-Aviv 69978, ISRAEL.
}

\end{center}

\begin{abstract}

We present the alternative topological twisting of $N=4$ \YM, in which
the path integral is dominated not by instantons, but by flat
connections of the {\it complexified\/} gauge group.  The theory is
nontrivial on compact orientable four-manifolds with nonpositive Euler
number, which are necessarily not simply connected.  On such manifolds,
one finds a single topological invariant, analogous to
the Casson invariant of three-manifolds.

\end{abstract}

\end{titlepage}

\newpage

\section{Introduction}

One of the simplest ways of making a topological field theory is to take a
theory with an extended spacetime supersymmetry in Euclidean space
and to ``twist it'',
changing the spins of the fields~\cite{wit:don,twisted}.  The resulting
theory will be topological if the twisted supersymmetry generators
include at least one spacetime scalar, which can then be thought of as a
BRST operator.  The prototypical example of this is Witten's twisting of
pure \ntwo{} \YM{}~\cite{wit:don}.  \ntwo{} supersymmetry has a \utwo{}
R--parity group; the twisting involves identifying its \sutwo{} with one of
the \sutwo's of the spacetime Lorentz group, and interpreting its
\uone{} as an anomalous ``ghost number'' symmetry.  Under this twisting,
one of the fermions of the theory gives rise to a self-dual antisymmetric
2--tensor $\chi_\munu^+$,
whose BRST transformation is the self-dual part of the \YM{} field
strength $F_\munu^+$.  Since topological theories are controlled by fixed
points of their BRST transformations~\cite{wit:n-matrix},  the path
integral is dominated by (anti)-instanton configurations.
Witten studied the cohomology of this theory, and showed that the
amplitudes of these operators calculate the Donaldson invariants of
the underlying manifold~\cite{donaldson}.

In the \ntwo{} case the twisting procedure is unique, so the most general
topological theory is formed by twisting \ntwo{} \YM{} coupled to \ntwo{}
matter hypermultiplets in various representations of the gauge
group~\cite{jon,n=2matter}.  Such theories have the same cohomology as the
Witten--Donaldson (WD) theory, but their ghost-number anomalies, ground
states, and amplitudes are changed.  To have a fundamentally
different theory, one needs to increase the number of supersymmetries.
Since topological gravity does not come from twisting supergravity
theories, the only possibility is to start from \nfour{} \YM.  This theory is
uniquely specified by its gauge group, but its  \sufour{} R--parity group
can be twisted in three different ways, leading to three distinct
topological theories~\cite{wit:s-dual}.

The lagrangians of the first
two theories were written down by Yamron, using superspace
techniques~\cite{jon}.  The first---which we shall call the ``half-twisted''
theory---involves simply treating the \nfour{} \YM{} as an \ntwo{} theory.
It is thus a particular (but the first, and maybe an especially
interesting) example of a WD theory with matter.  Explicitly, the
twisting is carried out by breaking the \sufour{} to an \sutwotwoone, with
the $\four \to (2,1)^1 \oplus (1,2)^{-1}$, and identifying one of
the \sutwo's with a spacetime \sutwo.  The other \sutwo{} remains
an internal symmetry of the theory.

The second twisting---giving rise to what
we shall call the Yamron--Vafa--Witten
(YVW) theory---is found by reducing the \sufour{} to an \sofour{},
with the $\four \to \four$, and identifying one of the resulting \sutwo's with
a spacetime \sutwo.  This theory thus has an ``\sutwo{} ghost-number''
symmetry, and {\it two\/} BRST symmetries, transforming as a doublet
of this \sutwo~\cite{jon}.  There is also a doublet of $\chi_\munu^+$ fields,
which both transform into $F_\munu^+$, so the theory is again dominated by
instantons.  Because such a nonabelian ghost number can not be
anomalous, the partition function is nonvanishing, and
indeed is the only observable of the theory.
Vafa and Witten showed that under appropriate conditions it gives the
Euler number of the moduli space of instantons, and that it
transforms covariantly under certain
S--duality transformations~\cite{wit:s-dual}.

The existence of the third twisting was pointed out in~\cite{jon}, as
a private communication from E.~Witten, and again in~\cite{wit:s-dual}.
This theory has not been explicitly constructed, and it shall be the
interest of this paper.  It can be obtained by further twisting the
internal \sutwo{} of the half-twisted theory\footnote{Thus
justifying our nomenclature.  In some ways, the three theories are
analogous to the half-twisted, A and B \ntwo{} topological sigma models in
two dimensions~\cite{B}.} with the remaining spacetime \sutwo, and
differs basically from the \ntwo{} and the YVW theories.  The
theory has a \uone{} ghost-number symmetry, and two BRST
operators, both with ghost number $+1$.  These are interchanged
by a hermitian conjugation operation, unlike in the YVW theory, where the two
BRST charges are equivalent.

The fields of the twisted theory are:
\vskip 3 mm
\begin{center}
\renewcommand{\arraystretch}{1.2}
\begin{tabular}{|c@{\ \ \ $\to$\ \ \ }c|c|c|}
\hline
\strut
\nfour{} YM                & twisted theory               & dimension
                                                               & ghost number
\\ \hline
$A_\mu$                    & $A_\mu$                      & 1
                                                               & $\phantom-0$
\\ \hline
$\Psi^I , \; \Psi_I^{(c)}$ & $\chi_\munu$                 & $3/2$      & $-1$
\\ \hline
                           & $\psi_\mu , \; \psibar_\mu$  & $3/2$
                                                               & $\phantom-1$
\\ \hline
                           & $\eta , \; \etabar$          & $3/2$      & $-1$
\\ \hline
$\Phi_{IJ}$                & $\V_\mu$                     & 1
                                                               & $\phantom-0$
\\ \hline
                           & $B$                          & 1
                                                               & $\phantom-2$
\\ \hline
                           & $C$                          & 1          & $-2$
\\ \hline
nothing                    & $P$                          & 2
                                                               & $\phantom-0$
\\ \hline
\end{tabular}
\end{center}
\vskip 2 mm
They are all in the adjoint representation of the gauge group.
(The twisted anticommuting fields actually
correspond to sums and differences of $\Psi^I$ and $\Psi_I^{(c)}$.  For
example, $\Psi^I$ twists to give the anti-self-dual part of $\chi_\munu$.)
Inspired by the half-twisted theory~\cite{jon}, we have introduced an auxiliary
field $P$, in order to help close the BRST algebra\footnote{One would also have
expected to have at least one auxiliary antisymmetric tensor field $B_\munu$
in analogy with the other twistings~\cite{jon}.
We have been unable to find the appropriate
auxiliary fields in our case, and this will cause some technical
nuisances later in the paper.}.   As in the WD~\cite{wit:don,jon} theories,
one can redefine a ``dimension'' to be the dimension minus half the
ghost number.  This results in dimensionless BRST charges,
but has little practical advantage.

There are several points worth noting about this spectrum:

\emptyliststuff{It is not chiral.}

\liststuff{Hermitian-conjugation:}
One can define an antihermitian
``hermitian-conjugation'' transformation, coming from the CP
invariance of the original \nfour{} \YM.  Under this conjugation
$\psi_\mu$ and $\eta$ are complex, transforming into $\psibar_\mu$ and
$\etabar$; $A_\mu$, $\V_\mu$, $B$, $C$, $P$ and the self-dual
piece of $\chi_\munu$ are real, and
the anti-self-dual piece of $\chi_\munu$ is purely imaginary:
\beqal
( A_\mu , \, \V_\mu , \, B , \, C , \, P ) & \; \to \;
        - \, ( A_\mu , \, \V_\mu , \, B , \, C , \, P ) \\
( \psi_\mu , \, \eta ) & \; \to \; - \, ( \psibar_\mu , \, \etabar ) \\
\chi_\munu  & \; \to \; - \, \chistar_\munu \stop \label{conj}
\eeqal
(Here $*$ denotes the Hodge dual:
$\chistar_\munu \equiv 1/2 \, \epsilon_{\mu\nu\rho\sigma}
\chi^{\rho\sigma}$; $\u{**}\n1\chi_\munu = \chi_\munu$.
The minus signs are because all the fields are in the adjoint
representation of the gauge group, represented by antihermitian matrices.)

\liststuff{$\chi_\munu$ is not self dual.}
Since $\chi_\munu$ is not self dual, one might
expect its BRST transformation to give the full $F_\munu$, so that
the path integral would be dominated by flat gauge-field configurations.
In fact, this is complicated by the fact that

\liststuff{There are 2 vector fields:}
We shall see that it is natural to combine $A_\mu$ and $\V_\mu$ into a
complex vector field $\A_\mu$:
\beqal
\A_\mu & \equiv A_\mu + i \, \V_\mu \\
\Abar_\mu & \equiv A_\mu - i \, \V_\mu \comma \label{A}
\eeqal
and we shall define three different covariant derivatives and field strengths:
\beqaltwo
D_\mu X & \equiv \pa_\mu X + \, \com {A_\mu} X \comma &
        F_\munu & \equiv \com { D_\mu} {D_\nu} \nonumber \\
\D_\mu X & \equiv \pa_\mu X + \, \com {\A_\mu} X \comma &
        \F_\munu & \equiv \com { \D_\mu} {\D_\nu} \\
\Dbar_\mu X & \equiv \pa_\mu X + \, \com {\Abar_\mu} X  \comma &
        \Fbar_\munu &\equiv \com { \Dbar_\mu} {\Dbar_\nu} \nonumber \stop
\eeqaltwo
Of course only $A_\mu$ is a true connection,
since the theory does not have a complexified gauge invariance.
Nevertheless, it will turn out that the path integral is dominated by flat
complexified connections $\A_\mu$, and the complexified gauge group will
play an important role in understanding the ground states of the theory.

\section{BRST transformations}

In principle, the BRST transformations of the theory could be found by
twisting the
supersymmetry transformations of the original \nfour{} \YM.  Instead, we use
these transformations and those of the half-twisted theory~\cite{jon} only as a
guide, and fix the transformations by demanding closure
of the algebra.  For convenience, we define the combinations
\beq
P^\pm \equiv P \pm i \, [ B \,,\, C ] \stop  \label{p-def}
\eeq
The $Q$ BRST transformations are given by:
\beqaltwo
& \Q \A_\mu = \twoi \, \psi_\mu &
                & \Q \Abar_\mu = 0 \\
& \Q \psi_\mu = 0 &
                & \Q \psibar_\mu = \Dbar_\mu B \\
& \Q C = i \, \eta &
                & \Q B = 0 \yesnumber \label{Q}\\
& \Q \eta = 0 &
                & \Q \etabar = - i \, P^- \\
& \Q P^- = 0 &
                & \Q \chi_\munu = \Fbar_\munu \semi
\eeqaltwo
they are clearly nilpotent, with $Q^2=0$.

Taking the conjugate of the $Q$ transformations with
the hermitian conjugation \noeq{conj},
one finds the $\Qbar$ transformations:
\beqaltwo
& \Qbar \Abar_\mu = \twoi \, \psibar_\mu &
                & \Qbar \A_\mu = 0 \\
& \Qbar \psibar_\mu = 0 &
                & \Qbar \psi_\mu = \D_\mu B \\
& \Qbar C = i \, \etabar &
                & \Qbar B = 0 \yesnumber \label{Qbar} \\
& \Qbar \etabar = 0 &
                & \Qbar \eta = i \, P^+ \\
& \Qbar P^+ = 0 &
                & \Qbar \chi_\munu = \Fstar_\munu \stop
\eeqaltwo
These are also nilpotent, with $\Qbar^2=0$.  This is unlike the
WD~\cite{wit:don} and YVW~\cite{jon} theories, in which the BRST
transformations close only
up to a gauge transformation.   However
the commutator of $Q$ with $\Qbar$ does give a
gauge transformation with parameter $\twoi B$:
\beq
\anticom Q \Qbar = T_{\twoi \, B} \stop
\eeq
This commutator also closes on $\chi_\munu$ only with the use of its equation
of motion.  This is a consequence of our inability to find appropriate
$B_\munu$ auxiliary fields, as we mentioned.

One sees that, as promised, the BRST transforms of $\chi_\munu$ give
rise to the complexified field strength $\F_\munu$, suggesting that the
theory is dominated by flat complexified connections.  This theory
reduces to the Witten--Donaldson theory if one demands reality under the
conjugation \noeq{conj}, and takes the BRST operator to be $Q+\Qbar$.
Then $\A_\mu$ becomes real,
only the self-dual part of $\chi_\munu$ survives, and the theory
reduces to a theory of instantons.

\section{The lagrangian}

Knowing the BRST transformations, one can now write the most general
renormalizable lagrangian with zero ghost number that is invariant
under both\footnote{If we were to
have demanded invariance just under $Q$, and not also under $\Qbar$,
the only extra generality that we would have had would have been the
freedom to have altered two of the coefficients in the four terms of $\call_1$
in \noeq{l1}.  (Recall from \noeq{p-def} that $P^+$ contains two terms.)
We shall not use this extra freedom, since such a change would be ugly
and, being BRST exact, would not affect any of our results.} $Q$ and $\Qbar$.
As in the untwisted \nfour{} theory, the
lagrangian depends on a coupling constant $g$ and a theta angle\footnote{One
might naively have expected to have three independent
theta angles, corresponding, say, to $\int F \wedge F$,
$\int \F \wedge \F$ and $\int \Fbar \wedge \Fbar$.  However, since
only $A$ is a true gauge field, all of these define the same
characteristic class: since $F \wedge F$ is locally the exterior derivative
of a \CS{} term, and
the difference between the \CS{} terms of $A$ and of $\A$ is
globally defined, all three terms are equal.} $\theta$.
After using ones freedom
to rescale fields, while preserving the two BRST transformations of
\noeq{Q} and \noeq{Qbar}, one sees that the most general lagrangian
depends upon only one more parameter $\alpha$.  It can be written as the
sum of three terms:
\beqalno
\call_1 & = \frac 1 {g^2} \; Q \Qbar \,
                        \Tr \left ( 2 \, \D_\mu C \, \V^\mu
                        + \alpha \, C \, P^+ \right ) \nonumber \\
        & = \frac 1 {g^2} \; Q \;
                        \Tr \left ( - 2 \, \D_\mu C \, \psibar^\mu
                                    + \twoi \, \D_\mu \etabar \, \V^\mu
                                    + i \, \alpha \, \etabar \, P^+ \right )
                                    \semi \label{l1} \\
\call_2 & = Q \; \Tr \left (
        - \, \frac 1 {2 \p1 g^2} \; \chi_\munu \, \F^\munu
        + \frac \theta {16\, \pi^2} \; \chistar_\munu \, \Fbar^\munu \right )
                                            \semi \label{l2} \\
\call_3 & = \frac {\twoi} {g^2} \; \Tr \left (
        \chistar^\munu \, \Dbar_\mu \psibar_\nu
        - \frac14 \, B \anticom { \chistar^\munu }  { \chi_\munu }
        \right )
                                            \stop \label{l3}
\eeqalno
Evaluating the BRST transformations, one finds:
\beqal
\call = \; \frac1{g^2} \; \Tr \;  \Bigl [ \, &
                - \,\frac12 \; \F_\munu \Fbar^\munu
            + \alpha \, \left ( P - \frac1\alpha \, D_\mu \, \V^\mu \right)^2
            - \frac1\alpha \, \bigl( D_\mu \, \V^\mu \bigr)^2  \\
        & - \left( \D_\mu C \; \Dbar^\mu B
            + \Dbar_\mu C \; \D^\mu B \right)
            + \alpha \, { \com B C }^2 \\
        &   + \twoi \, \psibar^\mu \, \D_\mu \eta
            + \twoi \, \psi^\mu \, \Dbar_\mu \etabar
            + \twoi \, \chi^\munu \, \D_\mu \psi_\nu
            + \twoi \, \chistar^\munu \, \Dbar_\mu \psibar_\nu \\
        &   - \twoi \, \alpha \, B \, \anticom \eta \etabar
            + \fouri \, C \, \anticom {\psi_\mu}{\psibar^\mu}
            - \, \frac i 2 \, B \anticom {\chi_\munu}{\chistar^\munu}
            \, \Bigr ] \\
        + \; \frac{\theta}{16\pi^2}  & \; \Tr \;
            \bigl ( \, \Fbar_\munu \Fstarbar^\munu \, \bigr ) \label{l} \stop
\eeqal
(Here we have integrated by parts to isolate the equation of motion of the
auxiliary field $P$.)
The untwisted \nfour{} theory
corresponds to $\alpha=1$, in which case the kinetic term of the $\V_\mu$'s is
simply proportional to $\V_\mu \, \Box \; \V^\mu$.
Important features of the lagrangian include:

\liststuff{$Q$ invariance:}
While $\call_1$ and $\call_2$ are manifestly exact under $Q$, $\call_3$
is not exact.  However, by using the Jacobi identity and the definition
of the field strength, one easily sees that $\call_3$, and therefore
$\call$, is invariant under $Q$.

\liststuff{$\Qbar$ invariance:}
It is less clear that $\call$ is invariant under $\Qbar$.  However
$\call$ is invariant under $Q$, and is real with respect to the
hermitian conjugation operation \noeq{conj}, so it is also invariant
under $\Qbar$.

\liststuff{topological nature:}
$\call_3$ is not $Q$--exact, but since it is a four form, its integral is
independent of the metric.  Thus the full {\it stress tensor\/} of
the theory is $Q$--exact.  This is the definition of a topological field
theory, and is sufficient to guarantee that physical quantities have no
dependence on the metric of the underlying four-manifold\footnote{If one
had had the appropriate auxiliary fields, the entire lagrangian would
presumably have been $Q \Qbar$ exact.  Such a situation occurs
in the 2--dimensional B--twisted topological sigma model of~\cite{B},
which was written with auxiliary fields in~\cite{lamba}.}.

\liststuff{scale, but not conformal invariance:}
The stress tensor of the theory is $Q$--exact, but it does not vanish.
Therefore, as in the WD theory~\cite{wit:don} and in the other
\nfour{} twistings~\cite{jon}, the topological theory is {\it not\/}
conformally invariant.  However, after using the equations of motion,
the trace of the stress tensor becomes the divergence of a current:
\beq
T_\mu^\mu \sim \, \frac4{g^2} \; \pa_\mu \, \Tr \;  \Bigl [ \,
                C \, D^\mu B
                - \frac1\alpha \, \bigl( D \cdot \V \bigr)\, \V^\mu
                + i \, \eta \, \psibar^\mu
                + i \, \etabar \, \psi^\mu
        \, \Bigr ] \comma \label{trace}
\eeq
so one does have a global scale (or Weyl) invariance~\cite{wit:don}.
This is easy to understand without having to have done this calculation,
since BRST invariance means that one can not add curvature terms, mass
terms or cubic scalar couplings to the original scale-invariant \nfour{}
\YM.  The scale invariance holds at the full quantum level, since
\nfour{} \YM{} is finite.

\liststuff{coupling independence:}
In many ways, $\alpha$
appears like a gauge-fixing parameter in the lagrangian, suggesting
the gauge fixing of a complexified gauge symmetry.  Terms involving it are
$Q$--exact (\cf{} \noeq{l1}), and physical quantities should not depend on it.
Unlike the other topological \YM{} theories, the instanton-number term
is also $Q$--exact (\cf{} \noeq{l2}),
so there should also be no dependence on $\theta$.  (This
apparently surprising result is actually pretty reasonable when one realizes
that the path-integral is dominated by flat connections, which have zero
instanton number.) While it is less obvious, the same is also true of
the coupling constant
$g$.  In addition to its appearance in exact terms, $g$ appears only in
the factor in front of $\call_3$.   This dependence can be removed without
disturbing the $Q$ BRST transformation
by rescaling $\psibar$ and $B$ by a factor of $g^2$, and $C$ and
$\eta$ by $g^{-2}$.  We do not perform such a
rescaling, since it obscures the symmetry between $Q$ and $\Qbar$, but
its existence shows that physical quantities also should not depend on
$g$.  Once again, this conclusion should be obvious in a formulation with
the appropriate auxiliary fields.

We conclude that {\it any amplitude of the theory should be a pure number,
depending only
on the gauge group and the topology of the manifold.\/}

\section{Ground states: The geometry of the moduli space}

Since the theory is independent of
the gauge coupling, one can see what it studies by
examining it in the weak coupling
limit, where it is dominated by its ground states~\cite{wit:don}.
As in the WD theory, one first needs to
``Wick rotate'' $C \to - B^\dagger$.  (This
is incompatible with our hermiticity operation \noeq{conj}, but that
 will not cause us any problems.)  After integrating out the
auxiliary field $P$, the purely bosonic part of
the action can be written as the sum of positive semi-definite
terms, plus the topological theta term:
\beqal
\cals_{B} = \; \frac1{g^2} \; \Bigl ( \; &
                \frac12 \; \norm{\F_\munu}^2
                + \frac1\alpha \, \norm{D_\mu \, \V^\mu}^2  \\
        & + \norm{\Dbar^\mu B}^2 + \norm{\D^\mu B}^2
            + \alpha \, \norm{ \! \com B {B^\dagger} }^2 \; \Bigr ) \\
        + \; \frac{\theta}{16\pi^2}  & \; \int \Tr \;
            \bigl ( \, \Fbar_\munu \Fstarbar^\munu \, \bigr )
            \label{l-pos} \comma
\eeqal
where $||X||^2$ means the integral of $\Tr \, X X^\dagger$ over the manifold,
soaking up indices with the spacetime metric where needed.
As $g \to 0$, the path integral is clearly dominated by
configurations which are at the absolute minimum of the action, within
each topological class.  This is attained when
all the objects in all the norms in \noeq{l-pos} vanish identically.
Note that since such configurations are flat ($\F_\munu=0$), the theta
term does not contribute, as expected, so
$\cals_{B}=0$ at the ground states.
In a topological theory, one can formally argue that the
path integral is dominated by the fixed
points of the BRST transformation~\cite{wit:n-matrix}.  Examining the
the $Q$ and $\Qbar$ transformations in \noeq{Q} and \noeq{Qbar}, one
sees that this agrees with the condition that $\cals_{B}$
vanishes.

The first condition for minimizing the action,
$\F_\munu=0$, has a clear geometric meaning, but the meaning of the
other conditions may seem rather less transparent.
Following the logic of Vafa and Witten in the YVW twisting,
one can look for ``vanishing theorems'' to eliminate these~\cite{wit:s-dual}.
Thus the scalar field $B$ is a kind of bosonic partner of the ghosts used for
fixing the gauge invariance of the theory, and its treatment is very
similar to that in the WD and YVW  theories.  The only slight complication is
that there are two connections in our case.  Its minimization condition is
\beq
\Dbar_\mu B = \D_\mu B = \com B {B^\dagger} =0 \comma \label{B-eq}
\eeq
with the hermitian-conjugate equations for $B^\dagger$.  This means that
$B$ and $B^\dagger$ can be simultaneously diagonalized by a gauge
transformation, and that their $\D_\mu$ and $\Dbar_\mu$ derivatives both
vanish.  Any solution to \noeq{B-eq} with non-zero $B$ implies that the
connection $\A_\mu$ is reducible, and such configurations cause trouble,
since they imply that the moduli space is not compact.
In Donaldson theory one usually restricts
oneself to the gauge groups \sutwo{} and \sothree, so any reducible
connection is either trivial, or lies completely in a \uone{} subgroup.
The latter are then discarded by considering only manifolds with
vanishing first Betti number $b_1$.
In section~\ref{anomalies} we shall see that this theory is nontrivial
only on manifolds with $b_1 \ge 1$, so we shall have no
choice but to deal with these reducible connections.

Continuing in this way, one could also try to find a vanishing theorem for
$\V_\mu$ on appropriate manifolds, as
Vafa and Witten did for the antisymmetric tensor field $B_{i j}^+$ in
their twisting~\cite{wit:s-dual}.  This can be done because,
if one ``chooses the Feynman gauge'' $\alpha=1$ to simplify the
$\V_\mu$ kinetic term,
an integration by parts leads to
a miraculous cancellation of the terms in the lagrangian linear in
$F_\munu$.
Dropping the instanton number term, the action for $A_\mu$
and $\V_\mu$ becomes:
\beq
g^2 \, \cals_{A,\V} =
                \frac12 \; \norm{F_\munu}^2
                + \, \norm{D_\mu \, \V_\nu}^2
                + \, \frac12 \, \norm{ \! \com {\V_\mu} {\V_\nu} }^2
                - \int R_\munu \, \Tr \; \V^\mu \, \V^\nu
                                                         \stop \label{vecs}
\eeq
Recalling that $\V_\mu$ is {\it anti-}hermitian, one sees that if the
Ricci tensor of the manifold is strictly positive definite as a matrix,
then $\V_\mu$ must vanish  at a minimum of the action $\cals_{A,\V} =0$.
Note that applying the same logic to the \uone{} case leads to the well-known
result that harmonic 1--forms on such
manifolds must vanish, so they have $b_1 =0$.
If the Ricci tensor is positive {\it semi-}definite---in particular if it
vanishes---one sees that all the components of a non-zero $\V_\mu$
at a minimum of the action commute, and so can be simultaneously
diagonalized by a gauge transformation.  Since $\V_\mu$ is also covariantly
constant, the gauge field is again reducible.
Examining the \uone{} case, one now finds the very strong
constraint that any harmonic 1--form on such a manifold must also be
a Killing vector.  This means that either $b_1=0$ again, so
$\V_\mu$ must vanish, or that, at least
locally,  $\calm$ is simply the product of a circle with a lower
dimensional manifold.

However, since this theory has nothing to calculate on
manifolds with $b_1=0$, we shall need to understand the geometric meaning
of the full moduli space with non-zero $\V_\mu$,
defined by the two equations
\bsubeq \label{moduli}
\beq
\F_\munu = \Fbar_\munu = 0 \label{flat}
\eeq
and
\beq
D \cdot \V = \; \frac{i}2 \, \com {\D_\mu} {\Dbar^\mu} = 0 \stop \label{gc}
\eeq
\esubeq
We have suggested that in many ways $\A_\mu$ acts as the
connection of a complexified gauge group $\gc$.  While it is impossible
to properly implement this idea
in the action itself, it is very important in understanding the moduli
space of the theory.  The flatness condition \noeq{flat} is indeed
invariant under $\gc$ transformations.  Imposing $D \cdot \V = 0$ then
looks like a partial gauge fixing of $\gc$ to the ordinary
gauge group \g.   More precisely, our claim is that the space of irreducible
solutions of \noeq{moduli}, modulo $\g$, is equivalent to the space
of irreducible solutions to $\F_\munu = 0$, modulo $\gc$.  In other words, the
moduli space is the space of flat connections of the complexified gauge
group.

Trivially, any point in the moduli space \noeq{moduli} gives
a single point in the moduli space of flat complexified connections.  The
problem is to show that the $\gc$--orbit of any flat complexified
connection contains one and only one point on the original moduli
space.  This is very similar to the Gribov problem of
choosing the coulomb gauge for an ordinary gauge field, and, in order
to argue for the {\it uniqueness\/} of a solution to \noeq{gc}, one can
modify a technique used there.  Thus, consider the orbit of any (not
necessarily flat) connection under {\it complexified\/} gauge
transformations $\Omega \in \gc$, with
$\D_\mu^\Omega \equiv \Omega^{-1} \D_\mu \Omega$,
and define the real, positive semi-definite function
\beq
F[\Omega] = - \, \frac12 \; \Tr \int g^\munu \, \V_\mu^\Omega \, \V_\nu^\Omega
        \label{F}
\eeq
on that orbit.  For an infinitesimal $\Omega \to 1 + \omega$,
one finds
\beq
\delta F = i \; \Tr \int g^\munu \, \V_\mu \, D_\nu \, \omega^+  \comma
\eeq
where $\omega^+ \equiv (\omega + \omega^\dagger)/2$ is the part of
$\omega$ in the algebra of the coset $\gc/\g$.  Thus $F$ has an extremum
exactly
when \noeq{gc} is satisfied.  At second order one finds
\beq
\delta^2 F = \, \norm{ \D_\mu \, \omega^+ }^2 \comma
\eeq
so any such extremum is a strict
minimum (except for the pure gauge directions), if the connection is
irreducible.  This means that
$F[\Omega]$ is a strictly convex Morse function, and the minimum must
be unique\footnote{One can also show the {\it uniqueness\/} of any irreducible
solution directly (This proof is due to Ori Ganor):  Assume that one
has one solution $\D_\mu$ to \noeq{moduli}, and substitute
$\D_\mu^\Omega$ into the ``gauge condition'' \noeq{gc}.  Multiplying on
the left by $\Omega$, and on the right by $\Omega^\dagger$, one finds
$g^\munu \,  M \, \Dbar_\nu \left( M^{-1} \; \D_\mu M \right) = 0$,
where $M \equiv \Omega \, \Omega^\dagger$.
Integrating this over the manifold, and doing an integration by parts,
one sees that $\D_\mu M$ necessarily vanishes.  The crucial difference
from the Gribov case is that here $M^{-1}$ is a hermitian matrix.  The
irreducibility of the connection now implies that $M$ is in the trivial
representation of $\g$, and so is proportional to the unit matrix.  The
only possibility is $M = \Omega \, \Omega^\dagger =1$, meaning that
$\Omega$ is an ordinary $\g$ transformation.  Thus the solution to
\noeq{gc} is unique.}.  (This proof is a particular example of a result
of Kempf and Ness~\cite{kempf}.)

The problem is that, while it may be plausible, we have not
proven that the minimum of $F[\Omega]$ actually {\it exists}.
This can be done using
``moment-maps''\footnote{I would like to thank Jeremy Schiff for
introducing me to moment maps, and for
pointing out that $D \cdot \V$ acts like one.}.
(We actually
need a generalization of this idea to infinite-dimensional spaces, as
used, for example, in Atiyah and Bott's study of the moduli
space of \YM{} theories on a Riemann surface ~\cite{atiyah-bott}.)
First note that the space of complexified connections has a natural
symplectic structure, with $A_\mu$ and $\V_\nu$ satisfying the Poisson
bracket
\beq
\anticom {A_\mu} {\V_\nu} = g_\munu \stop
\eeq
Using this bracket, one easily sees that (ordinary) gauge
transformations are generated by the moment map $\mu(A,\V) = D \cdot \V$,
which is just our familiar ``gauge-fixing condition'' \noeq{gc}.
Guillemin and Sternberg showed that the quotient of
the full space by $\gc$ is equivalent to the quotient of $\mu^{-1}(0)$ by
$\g$~\cite{sternberg}.  In other words, \noeq{gc} uniquely fixes
$\gc$ to $\g$, and our moduli space is indeed
the space of flat complexified connections divided by complexified
gauge transformations.

\section{Operators, anomalies and amplitudes} \label{anomalies}

Now, in order to study the amplitudes in a topological field theory, one
needs to find the operators in the BRST cohomology, and
to determine the ghost-number anomaly of the theory.
In the WD theory the cohomology starts with ``instanton number'' 4--form
$F \wedge F$, with ghost number 0.  Following the usual procedure, this
descends to give a 3--form, with ghost number 1, and so on
until one reaches a 0--form with
ghost number 4~\cite{wit:don}.
The ghost-number anomaly depends on the instanton
number $k$, and the Euler number $\chi$ and signature $\sigma$ of the
manifold.  For \sutwo{} one has
\beq
\Delta  Q_{GH} \, (\hbox{WD}) =
  \, 8 k - \frac32 \, \bigl( \chi + \sigma \bigr)
   = 8 k - 3 \left( 1 - b_1 + b_2^+ \right) \stop
\eeq
This is generically nonvanishing, and it gives the (virtual) dimension of the
moduli space of instantons on the manifold.  The expectation value of sets
of operators---in practice of 2--forms---that balance the anomaly then give
the Donaldson invariants of the manifold~\cite{wit:don}.
One has the same cohomology, but a different anomaly,
in the WD theories with extra matter~\cite{jon,n=2matter}.

In the YVW theory, one needs to
arbitrarily pick one of the two BRST operators of the theory, say
$Q^1$.  Its cohomology is then the same as in the WD theory.  (These will
not be in the cohomology of $Q^2$.)   The ghost number is actually
part of an unbroken \sutwo, so it has no anomaly.  Since all the
operators in
cohomology have positive ghost number, the only nonvanishing amplitude
of the theory is its partition function.  Barring holomorphic
anomalies, this is a holomorphic function of $\tau \equiv \theta/2\pi$,
with interesting properties under modular
transformations of $\tau$, coming from S--duality~\cite{wit:s-dual}.

In our case one might think that the BRST cohomology would again be a
copy of the WD cohomology.   Indeed the WD ghost-number 4 scalar operator
\beq
O^{(0)} = \Tr  \; \bigl(  B^2  \, \bigr) \label{o0}
\eeq
is in the cohomology of both $Q$ and $\Qbar$.  However, here its descent
is somewhat unusual.  One first finds a 1--form in the
$Q$--cohomology, say, which is $\Qbar$ exact:
\beq
O^{(1)} = \Tr \; \bigl( B \, \psibar \, \bigr) =
       \Qbar \, \Tr \; \bigl( B \, \V \, \bigr) \label{o1} \stop
\eeq
This then descends to a 2--form
\beq
O^{(2)} = \Tr \; \bigl( B \, \Fstar -i \, \psibar \wedge \psibar \, \bigr) =
\Qbar \, \Tr \; \bigl( B \, \chi - i \, \V \, \psibar \, \bigr)
                        \label{o2} \comma
\eeq
which is not only $\Qbar$ exact, but which also closes only with the use of
the $\chi$ equation of motion.  The procedure then stops, and one does
not find any 3-- or 4--forms in the cohomology.  In particular, we already
know that the instanton number 4--form is exact.
We have not found any other operators in the BRST cohomology.

All the twisted \nfour{} theories can have at most purely
gravitational anomalies, since their ghost number symmetries are subgroups of
the \sufour{} R--parity of the original \nfour{} theory, which is anomaly
free.  Thus the most simple way to calculate the anomaly in this theory
would be to turn
off all the bosonic fields.  For the moment, we shall prefer to work
in the more general background of an irreducible flat connection.   The
covariantized exterior derivative $\D$ then squares to 0, so one has a
twisted elliptic complex
\beq
0 \onarrow{} \Omega^0 \onArrow{\D} \Omega^1 \onArrow{\D} \Omega^2
\onArrow{\D} \Omega^3 \onArrow{\D} \Omega^4 \onarrow{} 0  \comma
                            \label{complex}
\eeq
where $\Omega^n$ denotes the space of $n$--forms in the adjoint of the
complexified gauge group $\gc$.  Recalling that $B$ and $B^\dagger$ vanish in
these backgrounds, one can see that the zero modes of
$\eta$, $\psi$, $\chistar$, $\u * \psibar$ and $\u * \etabar$ are the harmonic
representatives of the cohomology classes $\H{n}$ of this complex, for
$n=0\ldots4$.  For example, by considering the
equations of motion involving $\psi$, one can see that it satisfies
\beq
\D \, \psi = \Dbar * \psi =0 \stop
\eeq
Because the ghost numbers of these fermions alternate between $-1$
and $1$, the anomaly is minus the
index of this complex.  To calculate this one can now turn $\A$ off, so
the index is simply the
dimension of $\g$ times the index of the de~Rham complex.  Thus\footnote{This
is the usual anomaly that one would
expect in a theory of flat connections~\cite{phys-rep}.  The only
surprising feature of our construction is that we need a complex with
complexified connections.  It is interesting that if
one tries to construct a topological field theory of flat connections on
$\g$ directly in four dimensions, as has been done in two and three
dimensions~\cite{cobi}, one is again forced to consider complexified
connections~\cite{cobi-priv}.}
\beq
\Delta  Q_{GH}  =   \, - \chi \dim \g
                = \dim \g \; \bigl ( 2 \, b_1 - 2 - b_2   \bigr )
                                               \stop \label{anomaly}
\eeq
This anomaly has dramatic consequences: Since all of the
operators in the BRST cohomology have strictly positive ghost number,
the theory can have nontrivial amplitudes only over
manifolds with zero or negative Euler number.  Such manifolds have
$b_1 \ge 1$, meaning that one will be forced to consider \uone--reducible
connections.

Since $O^{(1)}$ and $O^{(2)}$ are $\Qbar$--exact, and all the
operators in the $Q$ cohomology are closed under $\Qbar$,
any amplitude involving them will vanish.  Thus one needs only to
calculate amplitudes of $O^{(0)}$'s.  The explicit form of $O^{(0)}$
in \noeq{o0} is unique only for \sutwo{} or \sothree, which are the
groups usually studied in Donaldson
theory.  In those cases $O^{(0)}$ has ghost number 4, and
there is a unique nonvanishing amplitude\footnote{For \sothree{} one
needs to sum over bundles with all possible
second Stiefel--Whitney classes $w_2$~\cite{wit:s-dual}.  Since $\Omega$ is a
pure number, S--duality should imply that it is the same for \sutwo{}
and for \sothree.}
\beq
\Omega = \vev{ \; \prod_{i=1}^{- \chi \dim \g /4}
          \Tr  \; \bigl(  B^2(x_i)  \bigr) } \comma \label{amp}
\eeq
where the $x_i$'s are arbitrarily chosen points on the manifold.  Note,
however, that this amplitude exists only if the Euler number of the
manifold is divisible by 4.  To find a nontrivial invariant on other manifolds,
one needs to consider larger gauge groups;  there will then be several
$\Omega$'s corresponding to various combinations of the casimirs of the groups.
In this case, the theory has the additional
complication that there may also be {\it nonabelian\/} reducible connections.

\section{Calculating the invariant $\Omega$}

In order to see what $\Omega$ actually calculates, let us
assume for the moment that we are at a point at which
there are no fermionic zero modes.  (Thus we
are considering a manifold with vanishing Euler number, and $\Omega$
reduces to the partition function of the theory.)  The moduli space now
consists of isolated points, so one only needs to see
what each point contributes to the partition
function~\cite{wit:s-dual}.  Noting that $\A_\mu$ is the only
field with a nontrivial background in this case, we expand
\beq
\A_\mu \to \A_\mu + a_\mu \stop
\eeq
Now, in order to evaluate the path integral, one must fix the gauge.  Since
$\D_\mu$ and $\Dbar^\mu$ commute \noeq{gc}, there is a
unique ``scalar laplacian'' $\boxa \equiv \D_\mu \Dbar^\mu$ at any ground
state of the theory.    We thus choose an obvious generalization of the Lorentz
gauge-fixing\footnote{In terms of $A_\mu$ and $\V_\mu$, this is
$\calg = D_\mu \, \delta A^\mu + [ \V_\mu \, , \, \delta \V^\mu ]$.},
\beq
\calg = \, \frac12 \; \left( \Dbar_\mu \, a^\mu + \D_\mu \, \abar^\mu \right)
                \label{fixing} \comma
\eeq
which transforms as $\delta \, \calg = \boxa \, \lambda$ under an
infinitesimal gauge transformation $\lambda$.  We then follow
the usual Faddeev--Popov procedure (in Feynman gauge),
and get the semiclassical gauge-fixed lagrangian by
expanding to quadratic order in the quantum fields:
\beqal
g^2 \, \call_{\rm quad} \, \to \, \Tr \;  \Bigl [ \, &
                a_\mu \left( g^\munu \, D^\rho \Dbar_\rho +
                      \com {\Dbar^\mu} {\D^\nu} \right) \abar_\nu
            + 2 \, C \, \boxa \, B
            + 2 \, b \, \boxa \, c \\
        &   + \twoi \, \psibar^\mu \, \D_\mu \eta
            + \twoi \, \psi^\mu \, \Dbar_\mu \etabar
            + \twoi \, \chi^\munu \, \D_\mu \psi_\nu
            + \twoi \, \chistar^\munu \, \Dbar_\mu \psibar_\nu \; \Bigr ]
             \label{l-quad} \stop
\eeqal
Here $c$ is the (anti-commuting) ghost and $b$ the antighost.
Recall that this semiclassical approximation becomes exact as $g \to 0$.

The functional integral over $B$ and $C$ now simply cancels that
over $b$ and $c$.  The integral over $a_\mu$ and $\abar_\nu$ gives
$\det^{-1}(-\boxamunu)$, where $\boxamunu$ is the
hermitian negative-semi-definite ``vector laplacian''
$g^\munu \, D^\rho \, \Dbar_\rho + [ \Dbar^\mu \, , \, \D^\nu]$.
This laplacian has no zero-modes at an isolated point in the moduli space.
The integral over the fermionic fields gives the Pfaffian of
their kinetic operators, which is convenient studied by squaring the
operators, and then taking the square root.
$\psi$ and $\psibar$ satisfy the squared equations of motion:
\beqal
D^\rho \, \Dbar_\rho \, \psi_\mu +
      [ \Dbar^\mu \, , \, \D^\nu] \, \psi_\nu &=
      \boxamunu \, \psi_\nu = 0\\
D^\rho \, \Dbar_\rho \, \psibar_\mu +
      [ \D^\mu \, , \, \Dbar^\nu] \, \psibar_\nu &=
      \boxamunuT \, \psibar_\nu =0 \comma
\eeqal
where $T$ indicates the operator transpose.  Thus, under our
assumption that there are no fermionic zero modes, the square of the
Pfaffian is $\det^2 \boxamunu$,  and one sees explicitly that the
Pfaffian cancels the
bosonic determinant up to a sign, as must occur in a topological
theory.  Because each eigenvalue of the fermion kinetic
operators occurs twice, once for $\psi$ and once for $\psibar$,
the sign must be positive.  Thus, in this simple case
the partition function simply counts
the number of points in the moduli space of the theory.
This is like the YVW twisting,
and unlike the WD theory, in which the partition function becomes a
{\it signed\/} sum over the points in the moduli space~\cite{wit:s-dual}.

On a manifold with negative Euler number one will necessarily have
zero modes of $\psi_\mu$ and $\psibar_\mu$.  For simplicity, we shall still
restrict ourselves to irreducible connections, so there are no
$\eta$ or $\etabar$ zero modes, and we shall {\it assume\/} that
there are no $\chi_\munu$ zero modes\footnote{
We have not been able to find any vanishing theorem for $\chi_\munu$
in our case, since the terms linear in the field strength do not cancel
upon an integration by parts.  It will be important to know in what
cases such zero modes do or do not occur.}.  From the index theorem (see
\noeq{anomaly}), this means that there are
$(-\chi)\cdot \dim \g /2$ zero modes each of $\psi_\mu$ and
$\psibar_\mu$, so that the moduli space is a complex manifold of
dimension $(-\chi)\cdot \dim \g /2$.  The calculation of $\Omega$ could
be carried out straightforwardly, by noting that these zero modes are
soaked up by bringing
down cubic $C \psi \psibar$ terms from the lagrangian, and then using Wick's
theorem to get $B$--$C$ propagators, but a simpler
method is to note that one can use the equation of motion of $C$ in the
semiclassical approximation~\cite{phys-rep}.  Thus, one integrates
\beq
\boxa B + \twoi \, \anticom {\psi_\mu}{\psibar^\mu} = 0 \label{B}
\eeq
to find $B$, substitutes this into the definition of $\Omega$ in
\noeq{amp}, and integrates the result over the zero modes of $a_\mu$,
$\psi_\mu$ and their conjugates.  Only the zero-mode parts of $\psi_\mu$
and $\psibar_\mu$ in \noeq{B} contribute, and once one has integrated
over them, $\Omega$ becomes the integral of a $(-\chi) \cdot \dim \g$ form over
the moduli space of flat connections.

Of course, the procedure that we have described is rather formal, and we
should state some caveats: On any manifold where $\Omega$ does not
trivially vanish by ghost-number conservation, there will necessarily be
\uone--reducible connections; in addition, there may be points on the
moduli space where $\H2$ is nonvanishing.  The moduli space will be
rather singular at both these sorts of points, and one will have to deal
with these singularities in order for $\Omega$ to be well defined.  Note
also that the identification of the moduli space of the theory with that of
complexified flat connections fails when connections are reducible.  An
additional problem that is specific to this particular
theory, is that the moduli
space may be noncompact because of the non-compactness of the
complexified gauge group $\gc$.  Finally, one should check that
BRST--exact variations of $\Omega$ do not give rise to contributions from the
boundaries or from the singular regions of moduli space.  All of these
issues must be dealt with in order to make $\Omega$ well-defined.

\section{Discussion}

We have studied a twisting of \nfour{} \YM{} which leads to a new type of
topological field theory (TFT).  The theory is defined on any smooth orientable
compact four-manifold $\calm$, with any semisimple gauge group $\g$.
Unlike the other topological \YM{} theories, which are dominated by
instanton configurations, the path integral here is dominated by flat
connections $\A_\mu$ in the {\it complexified gauge group\/} $\gc$.
Because of the ghost-number anomaly of the theory, and the structure of
its BRST cohomology, the theory is trivial when the Euler number $\chi$
of the four--manifold $\calm$ is positive.  Otherwise it gives rise to a
single amplitude $\Omega$, defined in eq\eq{amp},
which is the integral of a $(-\chi) \cdot \dim \g$ form over the moduli space
of flat complexified connections on $\calm$.

The moduli space of flat connections on $\calm$---let us call it
$\calc$---depends only on the gauge group and the fundamental group $\pim$ of
$\calm$.  It is given by homeomorphisms of $\pim$ to $\g$ (given by
integrating Wilson lines of the connection over nontrivial cycles), modulo
conjugation (gauge transformations):
$\calc = Hom(\pim,\g)/\g$~\cite{phys-rep}.  Thus any well-defined
quantity on $\calc$ is obviously a topological invariant of the
manifold, and is even invariant under homomorphisms that are not
diffeomorphisms.  Finding such invariants is orthogonal to the Freedman
program of classifying four-manifolds~\cite{freedman} (and the Donaldson
program of classifying smooth four-manifolds~\cite{donaldson}), in which
$\calm$ is usually taken to be simply connected---If $\pim=0$, $\calc$
contains only the trivial connection.

This means that $\Omega$ is indeed  a topological invariant of
$\calm$.  However, it is natural to ask ``Why does one need to deal with the
complexified group?'', and ``What is so special about $\Omega$?''.  In
other words, ``What has this TFT given us?''.  If one knows the fundamental
group of a manifold then $\calc$ is completely determined, and there
is no need to try to solve for $\Omega$.  However, $\pim$
is notoriously difficult to find.  One could hope that since $\Omega$,
and $\Omega$ alone, comes from a TFT, it would be the simplest object
on $\calc$ that one could actually calculate.
This is not really possible using the straightforward techniques of
our paper, in which one first has to directly construct
$\calc$.  However, it may be worth noting that the
reformulation of the Donaldson theory in terms of a TFT did not
lead to practical methods for calculating invariants, until
the techniques of Seiberg and Witten~\cite{wit-nat} were used to study
that theory in its infra-red limit, where it reduces to a much simpler
abelian theory of ``monopoles''~\cite{wit:mon}.
This particular technique is not helpful in our theory,
since it is completely scale invariant
at the quantum level.  (This is also the case for the YVW theory that
calculates the Euler number of the moduli space of
instantons.)
Nevertheless, we can hope that new techniques, possibly coming from the close
association of \nfour{} \YM{} with string theory, will become available to
study this theory.  Even if this does not occur, one knows at the very
least that $\Omega$ will satisfy cutting relations, simply because it
comes from a TFT.  This is presumably not the case for other invariants
on $\calc$.  Thus one should be able to calculate the $\Omega$'s of
manifolds formed by surgery.

It is interesting to compare this situation with that in three
dimensions.   One can define the ``Casson invariant'' $\lambda$ on
oriented three-manifolds using a cutting procedure~\cite{casson}.
Taubes has shown that, at least on homology
three spheres\footnote{$\lambda$ was originally
defined only for homology three spheres in~\cite{casson}; this was extended to
rational homology three spheres (\ie{} three-manifolds with $b_1=0$)
in \cite{rational}.  On such manifolds the moduli
space ``generically'' consists of a discrete number of
points~\cite{3-hom}, and one evades the issue of how to deal with the
\uone--reducible connections.}, $\lambda$ is the Euler number of flat \sutwo{}
connections~\cite{taubes}, and Witten has pointed out that this is given
by the partition function of a certain three-dimensional topological field
theory~\cite{wit:2+1,cobi,blau:cas}.  There is an
obvious analogy between $\Omega$ and $\lambda$.  The Casson invariant has
been very useful in classifying topological three-manifolds,
and we  hope that $\Omega$ could be useful in the four-dimensional
case.

\vskip 1 cm

{\large \bf \noindent Acknowledgments}

I would like to thank Maxim Braverman, Jeremy Schiff, Ori Ganor and Michael
Farber for their patience in many long and helpful discussions.  I am also
grateful to Cobi Sonnenschein, Yaron Oz, Ben Svetitsky and Alyosha Morozov
for helpful comments.


\newpage
{
\small
\parskip= 0 pt plus 2 pt
\def\baselinestretch{1.}

}


\begin{thebibliography}{99}

\bibitem{wit:don}
        E.~Witten,
        \cmp117,88,353.

\bibitem{twisted}
        E.~Witten,
                \cmp118,88,411,
                \np340,90,281;\\
        T.~Eguchi and S.-K.~Yang,
                \mpl5,90,1693;\\
        R.~Brooks and D.~Kastor,
                \pl246,90,99.

\bibitem{wit:n-matrix}
        E.~Witten,
        \np371,92,191.

\bibitem{donaldson}
        S.~Donaldson,
        \jou{Topology}29,90,257.

\bibitem{jon}
        J.P.~Yamron,
        \pl213,88,325.

\bibitem{n=2matter}
        D.~Anselmi and P.~Fr\'e,
        \np404,93,288, \hep9211121;
        \np416,94,255, \hep9306080;
        \pl347,95,247, \hep9411205;\\
        M.~Alvarez and J.M.F.~Labastida,
        \pl315,93,251, \hep9305028;
        \np437,95,356, \hep9404115;
        {\it ``Non-abelian monopoles on four-manifolds'',\/}
        preprint US--FT--4/95, \hep9504010.
        \\
        S.~Hyun, J.~Park and J.-S.~Park,
        {\it ``Topological QCD$_{c}$'',\/}
        preprint YUMS--95--08, CALT--68--1985,  SWAT/67, \hep9503201.

\bibitem{wit:s-dual}
        C.~Vafa and E.~Witten,
        \hep9408074, \np431,94,3.

\bibitem{B}
        E.~Witten,
        {\it ``Mirror manifolds and topological field theory'',\/}
        in {\it ``Essays on Mirror Manifolds'',\/}
        ed. S.T.~Yau
        (International Press, 1992), \hep9112056.

\bibitem{lamba}
        J.M.F.~Labastida and M.~Mari\~no,
        \hep9405151, \pl333,94,386.

\bibitem{kempf}
        G.~Kempf and L.~Ness,
        in {\it ``Algebraic Geometry'',\/}
        \jou{Lect.\ Notes in Math.}732,79,233
        (Springer-Verlag, 1979).

\bibitem{atiyah-bott}
        M.F.~Atiyah and R.~Bott,
        \philos308,82,523.

\bibitem{sternberg}
        V.~Guillemin and S.~Sternberg,
        \jou{Invent.\ Math.}67,82,491; \ibid{} 515.

\bibitem{phys-rep}
        See, for example,
        D.~Birmingham, M.~Blau, M.~Rakowski and G.~Thompson,
        \prp209,91,129.

\bibitem{cobi}
        J.~Sonnenschein,
        \prd42,90,2080.

\bibitem{cobi-priv}
        J.~Sonnenschein,
        private communication.

\bibitem{freedman}
        M.H.~Freedman,
        \jdiff17,82,357.

\bibitem{wit-nat}
        N.~Seiberg and E.~Witten,
        \hep9407087,
        \np426,94,19, erratum \ibid{} \noj430,94,485;
        \hep9408099, \np431,94,484.

\bibitem{wit:mon}
        E.~Witten,
        {\it ``Monopoles and four manifolds'',\/}
        IAS preprint IASSNS--HEP--94--96,
        \hep9411102.

\bibitem{casson}
        A.~Casson,
        {\it ``An invariant for homology 3--spheres'',\/}
        Lectures at MSRI, Berkeley (1985).

\bibitem{rational}
         K.~Walker,
         {\it ``An extension of Casson's invariant'',\/}
         (Princeton University Press, 1992).

\bibitem{3-hom}
        R.~Fintushel and R.J.~Stern,
        in {\it ``Geometry of low-dimensional manifolds: 1'',\/}
        ed. S.K.~Donaldson and C.B.~Thomas
        (Cambridge University Press, 1990).

\bibitem{taubes}
         C.H.~Taubes,
         \jdiff31,90,547.

\bibitem{wit:2+1}
        E.~Witten,
        \np323,89,113.

\bibitem{blau:cas}
        M.~Blau and G.~Thompson,
        \hep9112012, \cmp152,93,41.

\end{thebibliography}
\end{document}